\documentclass[aps,pra,twocolumn,reprint,floatfix,longbibliography]{revtex4}
\usepackage[utf8]{inputenc} 
\usepackage{graphicx}
\usepackage{amsmath}
\usepackage{amsfonts}
\usepackage{mathtools}
\usepackage{float}
\usepackage{tabularx, booktabs}
\usepackage{color}
\usepackage{braket}
\usepackage{bm}
\usepackage{comment}

\newcommand{\1}{\uparrow}
\newcommand{\2}{\downarrow}

\begin{document}
\title{Higher-order effective interactions for bosons near a two-body zero crossing} 
\author{A. Pricoupenko and D. S. Petrov}
\affiliation{Universit\'e Paris-Saclay, CNRS, LPTMS, 91405 Orsay, France}

\date{\today}

\begin{abstract}

We develop the perturbation theory for bosons interacting via a two-body potential $V$ of vanishing mean value. We find that the leading nonpairwise contribution to the energy emerges in the third order in $V$ and represents an effective three-body interaction, the sign of which in most cases (although not in general) is anticorrelated with the sign of the long-range tail of $V$. Explicit results are obtained for a few particular two-body interaction potentials and we perform a detailed perturbative analysis of tilted dipoles in quasi-low-dimensional geometries.

\end{abstract}

 \maketitle

\section{Introduction}

Systems with partially attractive and partially repulsive forces, fine-tuned to an approximate overall cancellation of the mean-field term, provide an interesting platform for studying various beyond-mean-field (BMF) phenomena, remarkable recent examples being quantum droplets and dipolar supersolids (see Ref.~\cite{Pfau} for review). In contrast to the mean-field energy, which is essentially the first-order Born integral of the interaction potential multiplied by the number of interacting pairs, the BMF term is sensitive to many-body effects reflecting the structure of the Bogoliubov vacuum, i.e., the spectrum of Bogoliubov quasiparticles and their density of states. This can lead to a rather exotic and nonanalytic dependence of the BMF energy density on the particle density ($n^{5/2}$ in three dimensions, $n^2\ln(n)$ for $D=2$, and $n^{3/2}$ for $D=1$). On the other hand, in quasi-low-dimensional regimes one can also recover the integer-power behavior with the leading terms in the energy density proportional to $n^2$ and $n^3$, which can be interpreted, respectively, as a renormalized two-body interaction and an emergent effective three-body force~\cite{Edler2017,Zin}. The latter has also been discussed for three elementary or composite bosons interacting with one another by a finite-range two-body interaction tuned to a zero crossing~\cite{Petrov2015,Grisha1D,Valiente,Grisha2D}. 

In this paper we reconcile the first-quantized few-body and the Bogoliubov perturbation theories in the particular case of a two-body potential of zero mean [defined by $\int V({\bf r})d^D r=0$ in the pure $D$-dimensional case and by Eq.~(\ref{TheAssumption}) in quasi-low-dimensional geometries] calculating the ground-state energy up to terms $\propto V^3$. We find that up to this order the result is an analytic function of the density and contains two-body corrections $\propto V^2n^2$ and $\propto V^3n^2$ as well as an effective three-body term $\propto V^3 n^3$. We present closed-form integral expressions for the corresponding coefficients in pure dimensions and in quasi-low-dimensional geometries and discuss their general consequences. We apply our formalism to bosons interacting by the double-Gaussian potential and by the Yukawa potential in pure dimensions. In particular, we establish that the emergent three-body interaction is repulsive (attractive) when the long-range tail of the underlying two-body potential is attractive (repulsive). We then calculate the three-body and two-body energy shifts for quasi-low-dimensional dipoles as a function of their tilt angle $\theta$ with respect to the confinement cylindrical symmetry axis. We find that the three-body force for quasi-two-dimensional dipoles changes from attraction to repulsion with increasing $\theta$. For one-dimensional dipoles the dominant three-body force is attractive and second order in $V$ except when they are aligned along the axis ($\theta=0$). In all these quasi-low-dimensional cases the confinement-induced shift of the two-body coupling constant is found to be positive as a result of a renormalization procedure.

The paper is organized as follows. In Sec.~\ref{Sec:PT} we use the standard perturbation theory to derive the interaction energy shift for $N$ atoms in free space and in quasi-low-dimensional geometries. In Sec.~\ref{app} we apply the obtained general formulas to the cases of double-Gaussian and Yukawa-plus-delta potentials in pure dimensions. Sections~\ref{Sec:2Ddipoles} and \ref{Sec:1Ddipoles} are devoted, respectively, to the quasi-two-dimensional and quasi-one-dimensional tilted dipoles. In Sec.~\ref{Sec:Bogoliubov} we make connections to the many-body case and show how our results can be obtained from the Bogoliubov theory. We conclude in Sec.~\ref{Sec:Conclusions}.

\section{Few-body perturbative approach\label{Sec:PT}}

We consider the system of $N$ distinguishable atoms~\cite{RemStat} characterized by the Hamiltonian (we assume unit mass and $\hbar=1$)
\begin{equation}\label{Ham}
\hat{H}=\sum_{i=1}^N-\partial^2_{{\bf x}_i}/2-\partial^2_{{\bf y}_i}/2 + U({\bf y}_i)+\sum_{i>j}V({\bf x}_i-{\bf x}_j,{\bf y}_i-{\bf y}_j),
\end{equation}
where $U({\bf y})$ is the confining potential and ${\bf x}$ and ${\bf y}$ denote the sets of single-particle coordinates in the unconfined and confined directions, respectively. For example, in the quasi-two-dimensional geometry ${\bf x}$ is the two-dimensional in-plane position vector and ${\bf y}=y$ is the coordinate perpendicular to the confinement direction. The unconfined space is assumed to be a cube of unit volume with periodic boundary conditions and we write the single-particle eigenstates of the noninteracting part of Eq.~(\ref{Ham}) as
\begin{equation}\label{OneBodyEigenstates}
\phi_{{\bf q},{\bm \nu}}({\bf x},{\bf y})=e^{i{\bf qx}}\psi_{\bm \nu}({\bf y}),
\end{equation}
where each component of ${\bf q}$ is an integer divided by $2\pi$ and ${\bm \nu}$ is the set of quantum numbers labeling the eigenstates $\psi_{\bm \nu}({\bf y})$ for the single-particle motion in the confined direction, $\epsilon_{\bm \nu}$ being the corresponding spectrum which we count relative to the ground state (such that $\epsilon_{\bm 0}=0$). 

Assuming weak interactions and applying the standard perturbation theory we write the ground-state energy of the $N$-body system as
\begin{equation}\label{SeriesEN}
E[N] = E^{(1)}[N] + E^{(2)}[N] + E^{(3)}[N] + ...,
\end{equation}
where $E^{(i)}[N]$ denotes the $i$-th order term in powers of the interaction strength. Here we distinguish the interaction strength, proportional to the typical amplitude of the potential $V({\bf x},{\bf y})$, from the two-body scattering amplitude, which could be fine-tuned to zero (see below). In Eq.~(\ref{SeriesEN}) we have already used the fact that by construction $E^{(0)}[N]=0$ (all particles are in the state $\{{\bm q},{\bm \nu}\}={\bf 0}$). We will restrict ourselves to perturbation order $i\leq 3$ and use the general expressions for the energy corrections available up to this order in Ref.~\cite{LL3}. 
Let us reproduce these general formulas for reference. Denoting the noninteracting multiparticle states by symbols with the bar $\bar{n}=\{{\bf k}_1,{\bm \nu}_1,...,{\bf k}_N,{\bm \nu}_N\}$,  the corresponding multiparticle energies by $\omega_{\bar{n}}$, their differences by $\omega_{\bar{n}\bar{m}}=\omega_{\bar{n}}-\omega_{\bar{m}}$, and the whole interacting part of the Hamiltonian (\ref{Ham}) by $\bar{V}$, energy corrections to state $\bar{n}$ read~\cite{LL3}
\begin{equation}\label{E1LL}
E_{\bar{n}}^{(1)}=\bar{V}_{\bar{n}\bar{n}},
\end{equation}  
\begin{equation}\label{E2LL}
E_{\bar{n}}^{(2)}=-{\sum}'_{\bar{m}} |\bar{V}_{\bar{m}\bar{n}}|^2/\omega_{\bar{m}\bar{n}},
\end{equation}  
\begin{equation}\label{E3LL}
E_{\bar{n}}^{(3)}={\sum}'_{\bar{k}} {\sum}'_{\bar{m}} \frac{\bar{V}_{\bar{n}\bar{m}}\bar{V}_{\bar{m}\bar{k}}\bar{V}_{\bar{k} \bar{n}}}{\omega_{\bar{m}\bar{n}}\omega_{\bar{k} \bar{n}}}-E_{\bar{n}}^{(1)}{\sum}'_{\bar{m}} \frac{|\bar{V}_{\bar{m}\bar{n}}|^2}{\omega_{\bar{m}\bar{n}}^2},
\end{equation}  
where the primes mean that the state $\bar{n}$ is excluded from the summations. 

Our task thus reduces to counting multi-particle excited states and calculating $\bar{V}_{\bar{n}\bar{m}}$. To this end we introduce the two-body matrix elements 
\begin{equation}
\begin{aligned}
&V_{{\bm \mu}{\bm \nu}}^{{\bm \zeta}{\bm \eta}} ({\bf k})=[V_{{\bm \nu}{\bm \mu}}^{{\bm \eta}{\bm \zeta}} (-{\bm k})]^*=V_{{\bm \zeta}{\bm \eta}}^{{\bm \mu}{\bm \nu}} (-{\bm k})\\
&= \! \int \! d {\bf y} d {\bf y}' d{\bf x}e^{i{\bf k}{\bf x}}V({\bf x},{\bf y}-{\bf y}')\psi^*_{{\bm \zeta}} ({\bf y}) \psi^*_{{\bm \mu}}({\bf y}') \psi_{{\bm \eta}}({\bf y}) \psi_{{\bm \nu}}({\bf y}'),\label{Vequa}
\end{aligned}
\end{equation}
where the equalities in the first line follow from $V({\bf r})=[V({\bm r})]^*=V(-{\bm r})$, assumed to be valid throughout the paper. In terms of these matrix elements the first correction to the $N$-body ground-state energy reads
\begin{equation}\label{E1}
E^{(1)}[N]=g_2^{(1)}\binom{N}{2}=V^{{\bf 0 0}}_{{\bf 0 0}}({\bm 0})\binom{N}{2}
\end{equation}
and the second one can be written as
\begin{equation}\label{E2}
E^{(2)}[N] = g_2^{(2)}\binom{N}{2}+g_3^{(2)}\binom{N}{3},
\end{equation}
where
\begin{equation}\label{g22}
g_2^{(2)}=-\sum_{{\bf k},{\bm \nu},{\bm \mu} } \frac{|V^{{\bf 0 \bm \nu}}_{{\bf 0} {\bm \mu}}({\bf k})|^2}{k^2 + \epsilon_{\bm \nu} + \epsilon_{\bm \mu}}
\end{equation}
and
\begin{equation}\label{g32}
g_3^{(2)}=-6\sum_{{\bm \nu }} \frac{|V^{{\bf 0 0}}_{{\bm 0} {\bm \nu}}({\bm 0})|^2}{\epsilon_{\bm \nu}}.
\end{equation}
In Eqs.~(\ref{g22}) and (\ref{g32}) the summations exclude terms with vanishing denominators [equivalent to the prime in Eq.~(\ref{E2LL})]. Equation~(\ref{g22}) is just the second-order interaction correction for a single pair. It corresponds to (virtual) excitations of two atoms which, in the first interaction episode, get excited into states $\{{\bm \nu},-{\bf k}\}$ and $\{{\bm \mu},{\bf k}\}$ and, in the second interaction episode, get back to their ground states. 

Equation~(\ref{g32}) represents an effective three-body attraction, which appears in confined geometries for weak two-body interactions of the usual type [for which, in particular, $V^{{\bf 0 0}}_{{\bf 0}{\bm \nu}}({\bf 0})\neq 0$]. It has been discussed in the context of quasi-one-dimensional~\cite{Muryshev2002,Mazets2008} and lattice bosons~\cite{Tiesinga}. It can also be obtained by solving the Gross-Pitaevskii equation for the condensate wave function~\cite{Muryshev2002}. Accordingly, this term is absent in pure dimensions (in our derivation this follows from the fact that there are no transversal excitations and thus no summation over ${\bm \nu}$). The emergence of this term in our first-quantization analysis can be understood by going back to Eq.~(\ref{g22}) and reconsidering virtual excitations where only one particle is promoted to ${\bm \nu}\neq 0$ (in this case ${\bf k}$ should vanish because of the momentum conservation in the unconfined directions). Then the amplitude $V^{{\bf 0 0}}_{{\bf 0} {\bm \nu}}({\bf 0})$ should be replaced by $(N-1)V^{{\bf 0 0}}_{{\bf 0} {\bm \nu}}({\bf 0})$, since the atom can be excited by interacting with $N-1$ other atoms, not just one as implied in Eq.~(\ref{g22}). In addition, in Eq.~(\ref{g22}) these special one-particle events are counted twice per pair, i.e., $N-2$ times too many. Equation~(\ref{g32}) is meant to compensate for these ``errors'' in Eq.~(\ref{g22}) when $N>2$. In Sec.~\ref{Sec:Bogoliubov} we present a many-body approach to this problem based on the second quantization, where Eq.~(\ref{g32}) emerges in a more natural manner.

Proceeding to the calculation of the third-order correction let us represent it as 
\begin{equation}\label{E3_N}
E^{(3)}[N]=g_3^{(3)}\binom{N}{3}+g_2^{(3)}\binom{N}{2}+\delta^{(3)}+\sigma^{(3)}.
\end{equation} 
In Eq.~(\ref{E3_N})
\begin{equation}\label{g33}
g_3^{(3)}=6\sum_{{\bf k},{\bm \nu}, {\bm \mu}, {\bm \eta}}\frac{V^{{\bm 0}{\bm \eta}}_{{\bm 0}{\bm \nu}}({\bf k})V_{{\bm \eta}{\bm 0}}^{{\bm 0}{\bm \mu}}({\bf k})V_{{\bm \mu}{\bm 0}}^{{\bm \nu}{\bm 0}}({\bf k})}{(k^2+\epsilon_{\bm \nu}+\epsilon_{\bm \eta})(k^2+\epsilon_{\bm \nu}+\epsilon_{\bm \mu})},
\end{equation}
where the summation extends to indices satisfying the constraint ${\bf k}\neq {\bf 0}\lor ({\bm \nu}\neq {\bf 0}\land{\bm \mu}\neq {\bf 0}\land{\bm \eta}\neq {\bf 0})$ ($\lor$ and $\land$ are boolean OR and AND, respectively). The term (\ref{g33}) accounts for the following sequence of virtual excitations of three different atoms. The first and the second atoms interact with each other and get excited into states $\{{\bm \nu},-{\bf k}\}$ and $\{{\bm \mu},{\bf k}\}$, respectively. Then, the second interaction episode results in the second atom getting back to the ground state and the third atom being excited to state $\{{\bm \eta},{\bf k}\}$. Finally, the first and the third atoms interact with each other both going down to the ground state. The constraint on the summation indices mentioned above is imposed in order to count in Eq.~(\ref{g33}) only genuine three-body events and not two-body or one-body ones, which we will now discuss.

We write the coefficient in the second term in the right-hand side of Eq.~(\ref{E3_N}) in the form
\begin{equation}\label{g23}
g_2^{(3)}=\sum_{{\bm \nu}, {\bm \mu}, {\bm \eta}, {\bm \zeta}, {\bf k}, {\bf q}}\frac{V_{{\bm 0}{\bm \eta}}^{{\bm 0}{\bm \zeta}}(- {\bf q})V_{{\bm \eta}{\bm \nu}}^{{\bm \zeta}{\bm \mu}}({\bf q}-{\bf k})V_{{\bm \nu}{\bm 0}}^{{\bm \mu}{\bm 0}}(\bf k)}{(k^2+\epsilon_{\bm \nu}+\epsilon_{\bm \mu})(q^2+\epsilon_{\bm \eta}+\epsilon_{\bm \zeta})},
\end{equation}
where the sum is constrained only by the requirement that the denominator do not vanish [equivalent to the primes in Eq.~(\ref{E3LL}) or, mathematically, to $({\bf k}\neq {\bf 0}\lor {\bm \nu}\neq {\bf 0}\lor {\bm \mu}\neq {\bf 0}) \land ({\bf q}\neq {\bf 0}\lor{\bm \eta}\neq {\bf 0}\lor{\bm \zeta}\neq {\bf 0})$]. Equation~(\ref{g23}) is nothing else than the first term in the right-hand side of the general formula Eq.~(\ref{E3LL}) calculated for a single pair of atoms. However, similarly to Eq.~(\ref{g22}), Eq.~(\ref{g23}) does not properly account for some two-body excitations when $N>2$. We will show that a higher-order compensation term is required [denoted by $\delta^{(3)}$ in Eq.~(\ref{E3_N})] which, however, vanishes when $V_{{\bm \mu}{\bm \nu}}^{{\bm \zeta}{\bm \eta}} ({\bf 0})=0$. Arguments for this are rather technical because of the chosen first-quantization technique. We present them for completeness in the next paragraph, which the reader can skip, if not interested.

Consider virtual-excitation trajectories of three particles in Eq.~(\ref{g23}), for which at least one of the atoms changes its state less than three times. In the sum of Eq.~(\ref{g23}) this happens when one of the matrix elements is of the form $V_{{\bm \alpha}{\bm \beta}}^{{\bm \gamma}{\bm \gamma}} ({\bf 0})$ or $V_{{\bm \gamma}{\bm \gamma}}^{{\bm \alpha}{\bm \beta}} ({\bf 0})$, i.e., we are dealing with an interaction event where only one atom changes its transversal state from ${\bm \beta}$ to ${\bm \alpha}$ leaving its momentum unchanged as well as the state of all other atoms. As one can see from Eq.~(\ref{g23}) this takes place when one or more of the following conditions is satisfied.
\begin{align}
&{\bm \nu}={\bf 0} \land {\bf k}={\bf 0},\label{a}\\
&{\bm \mu}={\bf 0} \land {\bf k}={\bf 0},\label{b} \\
&{\bm \eta}={\bm \nu} \land {\bf q}={\bf k},\label{c}\\
&{\bm \zeta}={\bm \mu} \land {\bf q}={\bf k},\label{d}\\
&{\bm \eta}={\bf 0} \land {\bf q}={\bf 0},\label{e}\\
&{\bm \zeta}={\bf 0} \land {\bf q}={\bf 0}\label{f}.
\end{align}
Then, the corresponding amplitude $V_{{\bm \alpha}{\bm \beta}}^{{\bm \gamma}{\bm \gamma}} ({\bf 0}) = V^{{\bm \alpha}{\bm \beta}}_{{\bm \gamma}{\bm \gamma}} ({\bf 0})$ in the sum of Eq.~(\ref{g23}) should be replaced by $V_{{\bm \alpha}{\bm \beta}}^{{\bm \gamma}{\bm \gamma}} ({\bf 0})+(N-2)V_{{\bm \alpha}{\bm \beta}}^{{\bm 0}{\bm 0}} ({\bf 0})$ in the case ${\bm \alpha} \neq {\bm \beta}$. The reason for this replacement is the ``wrong'' transition amplitude used in Eq.~(\ref{g23}); an atom can change its transversal state (from ${\bm \beta}$ to ${\bm \alpha}$) not only by interacting with the second atom in state ${\bm \gamma}$, but also with each of the other $N-2$ ground-state atoms. Similarly, for ${\bm \alpha} = {\bm \beta}$ [which can happen only for the combination (\ref{c})$\land$(\ref{d})] the matrix element $V_{{\bm \alpha}{\bm \alpha}}^{{\bm \gamma}{\bm \gamma}} ({\bf 0})$ should be replaced by the diagonal interaction matrix element for the multiparticle state $\{{\bf 0},...,{\bm \alpha},...,{\bm \gamma},...,{\bf 0}\}$ which incorporates interactions between all possible pairs (not only between the excited atoms). The replacement rule is $V_{{\bm \alpha}{\bm \alpha}}^{{\bm \gamma}{\bm \gamma}} ({\bf 0})\rightarrow V_{{\bm \alpha}{\bm \alpha}}^{{\bm \gamma}{\bm \gamma}} ({\bf 0})+(N-2)[V_{{\bm \alpha}{\bm \alpha}}^{{\bm 0}{\bm 0}} ({\bf 0})+V_{{\bm 0}{\bm 0}}^{{\bm \gamma}{\bm \gamma}} ({\bf 0})] + \binom{N-2}{2}V_{{\bm 0}{\bm 0}}^{{\bm 0}{\bm 0}} ({\bf 0})$. Finally, an additional modification is needed when only one atom is excited throughout the whole sequence of the three interaction events in Eq.~(\ref{g23}). This corresponds to combinations (\ref{b})$\land$(\ref{f}) or (\ref{a})$\land$(\ref{e}). In these cases not only the matrix elements should be corrected as explained above, but, in addition, the whole contribution of such terms should be divided by $N-1$ as in Eq.~(\ref{g23}) each of these ``one-body'' excitation sequences is counted twice for every pair [with subsequent multiplication by $\binom{N}{2}$ in Eq.~(\ref{E3_N})], whereas they should be counted only once per atom. These patches of Eq.~(\ref{g23}) can be cast in the form of a compensation term which we call $\delta^{(3)}$ but do not write explicitly as, for purposes of this paper, it is sufficient to understand that it vanishes when $V_{{\bm \mu}{\bm \nu}}^{{\bm \zeta}{\bm \eta}} ({\bf 0})=0$. 

Finally, the term $\sigma^{(3)}$ in Eq.~(\ref{E3_N}) corresponds to the last term in the general formula Eq.~(\ref{E3LL}). This term can be written explicitly by noting its resemblance to Eq.~(\ref{E2LL}). However, we observe that it is also proportional to Eq.~(\ref{E1}), which vanishes when $V_{{\bm 0}{\bm 0}}^{{\bm 0}{\bm 0}} ({\bf 0})=0$. 

In pure dimensions we have explicitly 
\begin{equation}\label{E3pure}
\begin{aligned}
&E^{(3)}[N]\\
&=\binom{N}{2}\sum_{{\bf q},{\bf k}}\frac{V(-{\bf q})V({\bf q}-{\bf k})V({\bf k})}{k^2q^2}+6\binom{N}{3}\sum_{{\bf k}}\frac{V^3({\bf k})}{k^4},
\end{aligned}
\end{equation}
where states with ${\bf k}={\bf 0}$, ${\bf q}={\bf 0}$ or ${\bf q}={\bf k}$ are excluded from the summation.

We observe that under the assumption  
\begin{equation}\label{TheAssumption}
\int   d{\bf x}V({\bf x},{\bf y})=0
\end{equation}
we have $V_{{\bm \mu}{\bm \nu}}^{{\bm \zeta}{\bm \eta}} ({\bf 0})\equiv 0$ for any set $\{{\bm \mu},{\bm \nu},{\bm \zeta},{\bm \eta}\}$ [see Eq.~(\ref{Vequa})]. Then, the energy of the $N$-body system, up to terms of order $V^3$, equals the renormalized two-body part $[g_2^{(2)}+g_2^{(3)}]\binom{N}{2}$ plus the leading nonpairwise part (an effective three-body interaction) given by $g_3^{(3)}\binom{N}{3}$. 

Note that the renormalized two-body interaction is asymptotically (for $V\rightarrow 0$) scales as $V^2$. Therefore, slightly softening the condition Eq.~(\ref{TheAssumption}) by allowing the Born integral be of order $\propto V^2$ gives us more flexibility in controlling the renormalized two-body interaction. In particular, one can make it weakly repulsive, weakly attractive, or vanishing. In the latter case the three-body term $g_3^{(3)}\propto V^3$ becomes the leading interaction correction as all other terms scale at least as $V^4$ [this follows from the scaling $V^{{\bm \nu}{\bm \mu}}_{{\bm \eta}{\bm \zeta}}({\bf 0})\propto V^2$ for the matrix elements at zero momenta].

This brings us to one of the main statements of our paper. In a sufficiently narrow vicinity of a two-body zero crossing, reached by a fine-tuned compensation of the attractive and repulsive parts of the interaction potential and conditioned by Eq.~(\ref{TheAssumption}), the dominant effective three-body interaction is third order in $V$ and is characterized by the coupling constant given by Eq.~(\ref{g33}). Note that this effective interaction can be repulsive or attractive depending on the shape of the two-body potential and on the confining geometry. In the next section we calculate this term explicitly for a few academically and practically relevant cases.

\section{Applications\label{app}}

\subsection{Double-Gaussian potential}

Gaussian potentials, although not very realistic, are very frequently used as model potentials for solving few-body and many-body problems. They are smooth, characterized by regular effective-range expansions, and allow one to make quite a few things analytically. Thus, the first example that we will consider is the sum of two Gaussians in pure dimensions, one attractive and one repulsive, with different ranges,
\begin{equation}\label{VGGx}
V(x)=v_0 e^{-\lambda_0 x^2} + v_1 e^{-\lambda_1 x^2}.    
\end{equation}

The condition (\ref{TheAssumption}) applied to (\ref{VGGx}) fixes the ratio $v_1/v_0$ as a function of $\alpha = \lambda_0 / \lambda_1$ and dimension $D$, namely $v_1/v_0 = -{\alpha}^{-D/2}$. Then, the potential (\ref{VGGx}) leads to the following three-body coupling constant in different dimensions
\begin{multline}
g_3^{(3)}(D=1) = \dfrac{3\pi}{2} \dfrac{{v_0}^3}{{\lambda_0}^3} \Big[\sqrt{3}(1-\alpha^{3/2}) \\  - {(2+ \alpha)}^{3/2} + {(1+2\alpha)}^{3/2}\Big],
\end{multline}

\begin{multline}
g_3^{(3)}(D=2) = \dfrac{9\pi^2}{8}  \dfrac{{v_0}^3}{{\lambda_0}^4} \Big[ - 2 \ln{(2+\alpha)}  - \alpha \ln{(3 \alpha^2  + 6 \alpha)} \\ + 2 \alpha \ln{(1+2 \alpha)} + \ln{(3+6\alpha)} \Big],
\end{multline}
and
\begin{multline}
g_3^{(3)}(D=3)   = \dfrac{3\pi^4}{4} \dfrac{{v_0}^3}{{\lambda_0}^5}  \Big[-\sqrt{3} + \sqrt{3\alpha} \\ + 3\sqrt{2+\alpha} - 3 \sqrt{1+2\alpha} \Big].
\end{multline}

In all these cases the configuration of the potential where it has a repulsive central part and attractive tail ($v_0>0$ and $\alpha>1$ or $v_0<0$ and $\alpha<1$) leads to a three-body repulsion. This phenomenon has been noticed in Ref.~\cite{Valiente}. In our perturbative analysis it follows from the fact that the Fourier transform of $V(x)$ is positive for any momentum. By contrast, the repulsive tail case leads to a three-body attraction.

Note also that the momentum integral in Eq.~(\ref{g33}) is converging at small momenta since $V(k)\propto k^2$ with the main contribution to the integral coming from momenta comparable to the inverse interaction range. We can thus say that the effective three-body term is characterized by the same range as the two-body potential.

\subsection{Yukawa-plus-delta potential}

We now consider the case of an attractive Yukawa potential compensated by a repulsive delta potential. A concrete realization of this model can be achieved by placing bosonic impurities (species $\2$) in a Bose-Einstein condensate of another species ($\1$). For simplicity, we assume that $m_\2=1\gg m_\1$ so that we can integrate out the host-gas dynamics in the adibatic Born-Oppenheimer approximation. In this manner the phonon exchange in the host gas leads to a static induced Yukawa attraction between the impurities and their direct interaction can be tuned in order to reach the neutrality of the total two-body force. This is attained at the phase-separation threshold $g_{\1\2}=\sqrt{g_{\1\1}g_{\2\2}}$, where $g_{\sigma\sigma'}$ are the two-body interaction coupling constants. In this case, the Fourier transform of the effective interaction (exchange plus direct) between two $\2$ impurities is given by
\begin{equation}\label{Yukawa}
V(k)=v-\frac{v/\xi^2}{k^2 + 1/\xi^2}=\frac{v k^2}{k^2 + 1/\xi^2},
\end{equation}
with $v = g_{\2\2}=g_{\1\2}^2/g_{\1\1}$ and $\xi = 1/\sqrt{4m_\1 g_{\1\1} n_\1}$ \cite{Stoof2000,Viverit2000}.

Substituting Eq.~(\ref{Yukawa}) into Eq.~(\ref{g33}) leads to the three-body coupling constant
\begin{equation}\label{g3Yukawa}
g_3 = S_Dv^3\xi^{4-D},
\end{equation}
where $S_1=3/8$, $S_2=3/(4\pi)$ and $S_3=9/(16\pi)$. Similarly to the double-Gaussian case, the attractive Yukawa tail corresponds to an effective three-body repulsion since $v$ and $V(k)$ are positive. One can also notice that the main contribution to the effective three-body interaction term comes from $k\sim 1/\xi$.

\subsection{Quasi-two-dimensional dipoles}\label{Sec:2Ddipoles}

We now consider quasi-two-dimensional dipoles in the geometry defined by ${\bf r}=\{x_1,x_2,y\}$, where ${\bf x}=\{x_1,x_2\}$ and $y$ are the in-plane and transverse coordinates, respectively. The external confinement potential is harmonic
\begin{equation}\label{U}
U(y)=y^2/2l^4 - 1/2l^2,
\end{equation}
where $l$ is the confinement oscillator length. The transversal eigenfunctions equal $\psi_\nu(y)=e^{-y^2/2l^2}H_\nu(y/l)/\sqrt{l\sqrt{\pi}2^\nu\nu !}$ and correspond to $\epsilon_\nu = \nu/l^2$. The dipole moments are assumed to be in the $\{x_1,y\}$ plane tilted by the angle $\theta$ with respect to the $y$-axis. The corresponding two-body interaction potential is the sum of the dipole-dipole and zero-range (pseudo)potentials~\cite{YiYou2000} 
\begin{equation}\label{dipolarV}
V({\bf r})=r_* \frac{r^2-3(x_1\sin\theta+y\cos\theta)^2}{r^5}+4\pi a \delta({\bf r})
\end{equation}
with the Fourier transform
\begin{equation}\label{FourierV3D}
V({\bf k},p)=4\pi r_*\left[\frac{(k_1\sin\theta+p\cos\theta)^2}{k^2+p^2}-\frac{1}{3}+\frac{a}{r_*}\right],
\end{equation}
where ${\bf k}=\{k_1,k_2\}$ and $k=|{\bf k}|$. Equation~(\ref{TheAssumption}) for this tilted-dipole setup translates to the condition~\cite{Fedorov2014,Baillie2015,Raghunandan2015}
\begin{equation}\label{aStarQuasi2D}
a=a_*=(1/3-\cos^2\theta)r_*,
\end{equation} 
which marks the point where  $V({\bf 0},p)=0$. 

Equation~(\ref{dipolarV}) should be understood as an effective mean-field potential valid in the limit of zero momenta and energies. Speaking about its two parameters, $r_*$ is imposed by the dipole moments and the short-range coupling constant $4\pi a$ is defined by postulating that the $N$-body interaction energy shift in the limit of extremely large $l$ scales as 
\begin{equation}\label{ENquasi2Dlimit}
E[N]=\binom{N}{2} \frac{1}{\sqrt{2\pi}l}4\pi(a-a_*).
\end{equation} 
This formulation of the zero-momentum limit avoids problems associated with the fact that this limit in Eq.~(\ref{FourierV3D}) is not well defined, which, in particular, makes Eq.~(\ref{E1}) useless in the strictly uniform three-dimensional space. 

The interesting for us matrix elements of the potential (\ref{FourierV3D}) can be written as~\cite{RemEdler}
\begin{widetext}
\begin{equation} \label{V2Dgeneral}
V_{\mu 0}^{\nu 0}({\bf k})=V_{\mu 0}^{0 \nu}({\bf k})=V_{0 \mu}^{0 \nu}({\bf k})=\int_{-\infty}^\infty \frac{dp}{2\pi}V({\bf k},p)\lambda_\nu(p)\lambda_\mu(-p),
\end{equation}
where 
\begin{equation}
\lambda_\nu(p)=\int_{-\infty}^\infty \psi_\nu(y) \psi_0(y) e^{ipy}dy=(-1)^{\nu/2}(lp)^\nu e^{-p^2l^2/4}/\sqrt{2^\nu \nu !}.
\end{equation} 
Integrating over $p$ in Eq.~(\ref{V2Dgeneral}) then gives
\begin{equation}\label{Vnu0mu0}
\begin{aligned}
&V_{\mu 0}^{\nu 0}({\bf k})=V_{\mu 0}^{0 \nu}({\bf k})=V_{0 \mu}^{0 \nu}({\bf k})=\frac{(-1)^{\mu+s/2}}{2^{(s-1)/2}}\sqrt{\frac{\pi}{\nu ! \mu !}}\frac{a-a_*}{l}[1+(-1)^s](s-1)!!-\frac{(-1)^{\mu+s/2}}{2^s}\sqrt{\frac{\pi}{\nu ! \mu !}}\frac{r_*}{l}(kl)^{s+1}e^{k^2l^2/2}\\
&\hspace{1.5cm}\times\left\{[1+(-1)^{s}](s-1)!!\,\Gamma\left(\frac{1-s}{2},\frac{k^2l^2}{2}\right)\left(\cos^2\theta-\frac{k_1^2}{k^2}\sin^2\theta\right)+\frac{1-(-1)^{s}}{\sqrt{2}}s!!\,\Gamma\left(-\frac{s}{2},\frac{k^2l^2}{2}\right)\frac{k_1}{k}\sin2\theta\right\}\\
&\hspace{5cm}\xrightarrow[s \gg 1]{}\sqrt{\frac{s!}{\nu ! \mu !}}\frac{1}{2^{s/2+5/4}\pi^{3/4}s^{1/4}l}[(-1)^{\mu+s/2} V({\bf k},\sqrt{s}/l)+(-1)^{\nu+s/2} V({\bf k},-\sqrt{s}/l)],
\end{aligned}
\end{equation}
\end{widetext}
where $\Gamma(j,\sigma)$ is the incomplete Gamma function and we have denoted $s=\nu+\mu$. The last line in Eq.~(\ref{Vnu0mu0}) is an approximate expression valid for large $s$ and obtained by observing that the product $\lambda_\nu(p)\lambda_\mu(-p)$ in this limit is essentially the sum of two delta peaks at $p=\pm \sqrt{s}/l$.

Substituting Eq.~(\ref{Vnu0mu0}) into Eq.~(\ref{g33}) and setting $a=a_*$ we obtain
\begin{equation}\label{g3TiltedDipoles}
g_3=\frac{{r_*}^3}{l}[C_0F_0(\theta)+C_1F_1(\theta)],
\end{equation}
where
\begin{equation}
F_0(\theta)=\frac{1+3\cos 2\theta}{4}\frac{31+12\cos 2\theta+21\cos 4\theta}{64},
\end{equation}
\begin{equation}
F_1(\theta)=\frac{1+7\cos 2\theta}{8}\sin^2 2\theta,
\end{equation}
and the numerical coefficients $C_0=-141.1$ and $C_1=-10.7$. The solid line in Fig.~\ref{Fig:g3} shows $C_0F_0(\theta)+C_1F_1(\theta)$ as a function of $\theta$. It equals $C_0$ for $\theta=0$, i.e., for dipoles aligned perpendicularly to the plane we arrive at a three-body attraction (assuming positive $r_*$). Again, we observe that the three-body attraction is correlated with the repulsive tail. By contrast, for dipoles in the plane $C_0F_0(\pi/2)+C_1F_1(\pi/2)=-5C_0/16$ and we predict a three-body repulsion.

That the sum in Eq.~(\ref{g33}) converges at large momenta and energies can be understood by considering the purely three-dimensional version of Eq.~(\ref{g33}) for dipoles. In this case, $V({\bf k})= O(|k|^0)$ and the integral over the three-dimensional momentum is converging at large $k$ as $\int d^3k/k^4$. A cutoff at $k\sim 1/r_0$ would produce an effective-range correction to $g_3$ on the order of $\delta g_3\sim r_0 r_*^3/l^2$~\cite{RemExtrapolation}. Assuming $r_0\sim r_*$, this gives the ``natural'' scaling $\delta g_3\sim {r_*}^4/l^2$ for the three-dimensional three-body interaction coupling constant $\propto r_*^4$ projected to the transversal ground state of the trap. This is to say that the three-body interaction (\ref{g3TiltedDipoles}) is enhanced by the factor $l/|r_*|\gg 1$ compared to the ``natural'' three-dimensional scale. Nevertheless, it remains much weaker than the ``natural'' two-dimensional scaling $g_3\propto l^2$ for a nonperturbative two-dimensional potential of range $l$.

The quasi-two-dimensional model at hand is often reduced to a purely two-dimensional one by projecting the interaction potential on the transversal Gaussian ground state~\cite{Fischer2006,Nath2009,Cai2010,Ticknor2011,Ticknor2012,Natu2013,Mulkerin2013,Fedorov2014,Raghunandan2015}. In our case this projection means that in Eq.~(\ref{g33}) we retain only terms with ${\bm \nu}={\bm \mu}={\bm \eta}={\bm 0}$. This approximation gives $C_0=-127.4$ and $C_1=0$ in Eq.~(\ref{g3TiltedDipoles}) and is rather accurate (see the dashed line in Fig.~\ref{Fig:g3}). This curious fact is consistent with the above-mentioned convergence of the sum in Eq.~(\ref{g33}) at high energies. As we will now show, the two-body energy correction requires a more accurate treatment.

\begin{center}
\begin{figure}[ht]
\vskip 0 pt \includegraphics[clip,width=1\columnwidth]{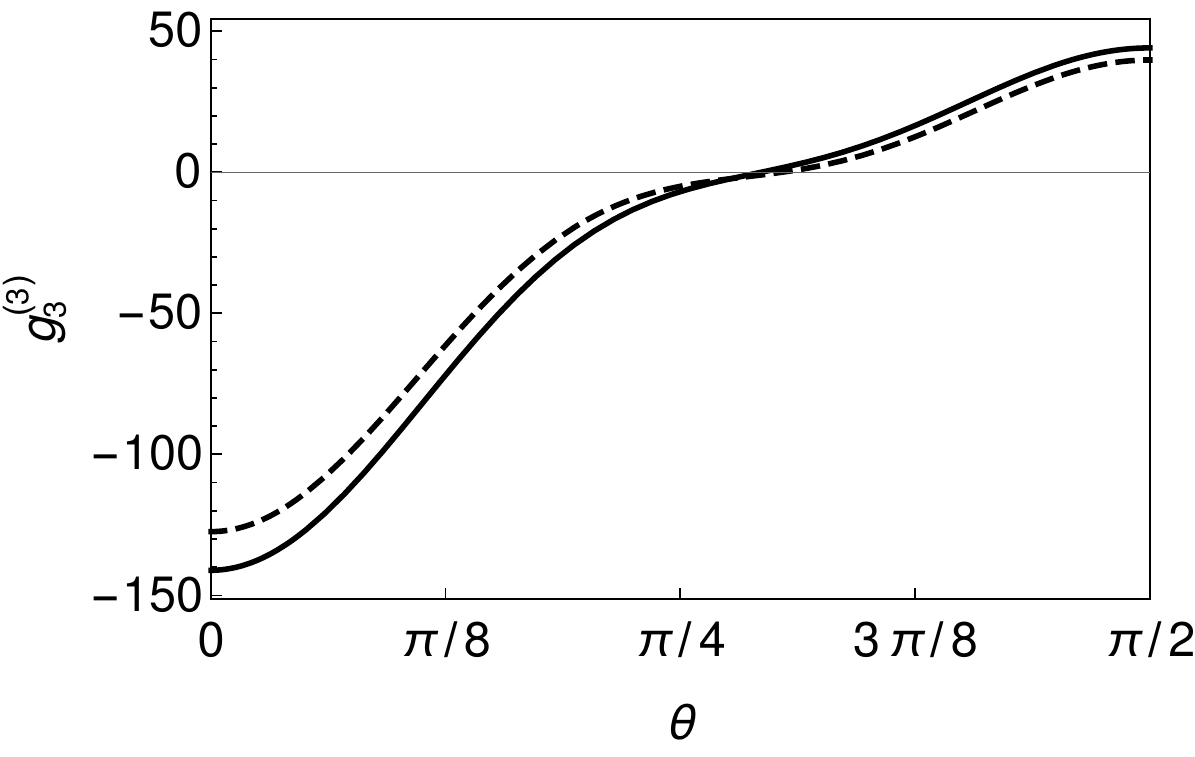}
\caption{
$g_3^{(3)}$ in units of ${r_*}^3/l$ as a function of the tilt angle $\theta$ for the case of quasi-two-dimensional dipoles. The solid line is obtained by evaluating the sum in Eq.~(\ref{g33}) with all excited states of the trap taken into account. The dashed line includes only the transverse ground state ($\nu=\mu=\eta=0$). Assuming $r_*>0$ the effective three-body interaction monotonically changes from attractive when dipoles are perpendicular to the plane ($\theta=0$) to repulsive when they are in the plane ($\theta=\pi/2$). The change of sign takes place at $\theta \approx 0.29\pi$. }
\label{Fig:g3}
\end{figure}
\par\end{center}

Let us now discuss the confinement-induced correction to the two-body interaction, given by Eq.~(\ref{g22}), for $a=a_*$. Under this condition the sum in Eq.~(\ref{g22}) converges at small momenta and we do not have to deal with the logarithmic infrared divergence, typical for the ``ordinary'' two-dimensional scattering. On the other hand, Eq.~(\ref{g22}) does feature an ultraviolet diverging part 
\begin{equation}\label{Sumasym}
-\frac{1}{\sqrt{2\pi}l}\sum_{s,{\bm k}}\frac{V^2({\bf k},\sqrt{s}/l)+V^2({\bf k},-\sqrt{s}/l)}{4\pi l\sqrt{s}(k^2+s/l^2)},
\end{equation}
which is obtained by substituting the large-$s$ asymptote given by the last line in Eq.~(\ref{Vnu0mu0}) into  Eq.~(\ref{g22}) and using the fact that $\sum_{\nu,\mu}(\nu ! \mu !)^{-1}\delta_{\nu+\mu,s}=2^s/s!$. By identifying $s=(lp)^2$ and passing from summation over $s$ to integration over $p$ Eq.~(\ref{Sumasym}) transforms into
\begin{equation}\label{2ndOrderBorn}
-\frac{1}{\sqrt{2\pi}l}\int_{-\infty}^\infty \frac{dp}{2\pi}\int \frac{d^2k}{(2\pi)^2}\frac{V^2({\bf k},p)}{k^2+p^2},
\end{equation}
which is nothing else than the second-order Born correction calculated for $V({\bf k},p)$ in free space and averaged over the transversal density 
profile~\cite{remLocal}. This piece renormalizes the short-range coupling constant $4\pi a$ and has to be formally thrown away since it has already been taken into account in Eq.~(\ref{ENquasi2Dlimit}). The regularized sum in Eq.~(\ref{g22}) for $a=a_*$ then equals
\begin{equation}\label{g2TiltedDipoles}
g_2^{(2)}=\frac{{r_*}^2}{l^2}\left(B_0\frac{3+10\cos 2\theta +19 \cos^2 2\theta}{32}+B_1\frac{\sin^2 2\theta}{4}\right),
\end{equation}
where the coefficients $B_0=0.55$ and $B_1=1.5$ are obtained by extrapolating the numerical summation to infinite cutoff. The second-order correction Eq.~(\ref{g2TiltedDipoles}) is positive (because of the renormalization) and monotonically decays from $B_0 (r_*/l)^2$ for $\theta=0$ (dipoles perpendicular to the plane) to $(3/8) B_0 (r_*/l)^2$ for $\theta = \pi/2$ (dipoles in the plane). This means that in order to stay at the two-body zero crossing while increasing the confinement, one has to tune the short-range interaction coupling constant to the value $4\pi a \approx 4\pi a_*-\sqrt{2\pi}g_2^{(2)}l$ (valid up to second order in $r_*/l$).

We should note that the positivity of the renormalized $g_2^{(2)}$ may be specific to the considered confinement and interaction potentials. Zin and co-workers~\cite{Zin}, using essentially the same renormalization scheme but for dipoles under periodic boundary conditions, arrived at a positive $g_2^{(2)}$.

\subsection{Quasi-one-dimensional dipoles}\label{Sec:1Ddipoles}

We now proceed to discussing the quasi-one-dimensional model of tilted dipoles, wich has recently been realized experimentally with Dy~\cite{Tang2018}. In spite of their formal analogy, the quasi-two-dimensional and quasi-one-dimensional models of tilted dipoles have an interesting difference which concerns the effective three-body interaction. Let us define the quasi-one-dimensional model by the coordinates ${\bf r}=\{x,{\bm y}\}=\{x,y_1,y_2\}$, the external confinement potential is assumed cylindrically symmetric, 
\begin{equation}\label{U1D}
U({\bf y})=y^2/2l^4 - 1/l^2.
\end{equation}
The single-particle eigenfunctions $\psi_{\bm \nu}({\bf y})$ satisfy $[-\partial^2_{\bf y}/2 + U({\bf y})]\psi_{\nu,m}=\epsilon_{\nu,m}\psi_{\nu,m}$, where $\epsilon_{\nu,m}=(2\nu+|m|)/l^2$. We have used the cylindrical symmetry of the potential (\ref{U1D}) to write ${\bm \nu}=\{\nu,m\}$, where the integers $\nu\geq 0$ and $-\infty<m<\infty$ are the radial and angular quantum numbers, respectively. The eigenfunctions read
\begin{equation}\label{EigenstatesQuasi1D}
\begin{aligned}
&\psi_{\nu,m}({\bm y})\\
&=e^{im\phi}\frac{(-1)^{\nu}}{l\sqrt{\pi}}\sqrt{\frac{\nu !}{(\nu+|m|)!}}\left(\frac{y}{l}\right)^{|m|} L_{\nu}^{|m|} \left(\frac{y^2}{l^2}\right)e^{-y^2/2l^2},
\end{aligned}
\end{equation}  
where  $L_{\nu}^{|m|}$ is the Laguerre polynomial and $\phi=\arg (y_1+iy_2)$.

The dipole moments are assumed to be in the $\{x,y_1\}$ plane tilted by the angle $\theta$ with respect to the longitudinal $x$-axis. The interaction (pseudo)potential is then given by
\begin{equation}\label{dipolarV1D}
V({\bf r})=r_* \frac{r^2-3(x\cos\theta+y_1\sin\theta)^2}{r^5}+4\pi a \delta({\bf r}),
\end{equation}
which can also be obtained from Eq.~(\ref{dipolarV}) by replacing $x_1\rightarrow y_1$, $x_2\rightarrow y_2$, and $y\rightarrow x$. Accordingly, the Fourier transform $V(k,{\bf p})$ of (\ref{dipolarV1D}) is obtained from Eq.~(\ref{FourierV3D}) by replacing $k_1\rightarrow p_1$, $k_2\rightarrow p_2$, and $p\rightarrow k$.

In contrast to the quasi-two-dimensional case, for quasi-one-dimensional dipoles with tilt, Eq.~(\ref{TheAssumption}) cannot in general be satisfied for all ${\bf y}$. Indeed, one can check that 
\begin{equation}\label{V01D}
\int dx V(x,{\bf y})=4\pi (a-a_*)\delta({\bf y})+2r_*\sin^2\theta\frac{y_2^2-y_1^2}{y^4},
\end{equation} 
where 
\begin{equation}\label{aeff1D}
a_*=\left(\frac{1}{3}-\frac{\sin^2\theta}{2}\right)r_*.
\end{equation}
From Eq.~(\ref{V01D}) we see that the condition $V_{{\bm 0}{\bm 0}}^{{\bm 0}{\bm 0}} ({\bf 0})=0$ requires $a=a_*$ (this condition corresponds to $\epsilon_{\rm dd}=1$ in notations of Ref.~\cite{Edler2017}). However, the matrix elements $V_{{\bm \mu}{\bm \nu}}^{{\bm \zeta}{\bm \eta}} ({\bf 0})$ involving excited states all vanish only if $a=a_*$ and $\sin\theta=0$. Therefore, for a finite tilt angle the dominant nonpairwise interaction correction is given by $g_3^{(2)}$ Eq.~(\ref{g32}) and corresponds to an effective three-body attraction. We will calculate it for arbitrary $\theta$. As far as $g_3^{(3)}$ is concerned, it can become dominant only at (or sufficiently close to) the point $\theta=0$.

The relevant matrix elements needed for these calculations can be written as~\cite{RemEdler}
\begin{widetext}
\begin{equation}\label{Vnu0mu01Dgen}
V_{{\bm 0} \{\mu,m'\} }^{{\bm 0} \{\nu,m\}}(k)=V_{\{\mu,-m'\} {\bm 0}}^{{\bm 0} \{\nu,m\}}(k)=V_{{\bm 0} \{\mu,m'\}}^{\{\nu,-m\} {\bm 0}}(k)=V_{{\bm 0} \{\mu,-m'\} }^{\{\nu,m\} {\bm 0}}(k)= \int \frac{d^2p}{(2\pi)^2}V(k,{\bf p})\lambda_{\nu,m}({\bm p})\lambda_{\mu,m'}(-{\bm p}),
\end{equation}
where
\begin{equation}
\lambda_{\nu,m}({\bm p})=\int\psi_0({\bm y})\psi_{\nu,m}({\bm y}) e^{i{\bm p}{\bm y}}d^2y=\frac{(-1)^{\nu+|m|/2}}{\sqrt{\nu !(\nu+|m|)!}}\left(\frac{lp}{2}\right)^{2\nu+|m|} e^{-p^2l^2/4} \left(\frac{p_1+ip_2}{p}\right)^m.
\end{equation}
Then, integrating in Eq.~(\ref{Vnu0mu01Dgen}) over ${\bm p}$ gives
\begin{equation}\label{Vnu0mu01D}
\begin{aligned}
&V_{{\bm 0} \{\mu,m'\} }^{{\bm 0} \{\nu,m\}}(k)=\frac{(-1)^{s}2^{-s-1}s!}{\sqrt{\nu ! (\nu +|m|)! \mu ! (\mu+|m'|)!}}\frac{4(a-a_*)\delta_{m+m',0}+r_*\sin^2\theta \sum_\pm \delta_{m+m',\pm 2}}{l^2}+\frac{r_*(-1)^s(kl/2)^{2s+2}e^{k^2l^2/2}}{l^2\sqrt{\nu ! (\nu +|m|)! \mu ! (\mu+|m'|)!}}\\
&\times \{s!\Gamma(-s,k^2l^2/2)[(4-6\sin^2\theta)\delta_{m+m',0}-\sin^2\theta\sum_\pm \delta_{m+m',\pm 2}]+2(s+1/2)!\Gamma(-s-1/2,k^2l^2/2)\sin2\theta \sum_\pm \delta_{m+m',\pm 1}\},
\end{aligned}
\end{equation}
\end{widetext}
where $s=\nu+\mu+(|m|+|m'|)/2$ and $\delta_{m,m'}$ is the Kronecker delta. 

Substituting Eq.~(\ref{Vnu0mu01D}) (more precisely, its particular case $V^{{\bm 0}{\bm 0}}_{{\bm 0} \{\nu,m\}}(0)=\frac{(-1)^{\nu}}{2^{{\nu}-1}}\left[\frac{a-a_*}{l^2}\delta_{m,0}-\frac{r_*\sin^2\theta}{8l^2}\sqrt{\frac{\nu+1}{\nu+2}}(\delta_{m,2}+\delta_{m,-2})\right]$) into Eq.~(\ref{g32})  we get
\begin{equation}\label{g31D}
g_3^{(2)}=-12\ln\frac{4}{3}\left(\frac{a-a_*}{l}\right)^2+\left(\frac{3}{2}-6\ln\frac{4}{3}\right)\left(\frac{r_*}{l}\right)^2\sin^4\theta.
\end{equation}
The first term in the right-hand side of Eq.~(\ref{g31D}) recovers the result of Refs.~\cite{Muryshev2002,Mazets2008} obtained for $r_*=0$. By tuning to $a=a_*$ this term vanishes simultaneously with the leading two-body energy shift $g_2^{(1)}=V^{{\bm 0}{\bm 0}}_{{\bm 0} {\bm 0}}(0)$. Nevertheless, Eq.~(\ref{g31D}) predicts that the three-body attraction persists for any finite tilt angle even if $a=a_*$. Tilted dipoles can thus realize a model of one-dimensional bosons with three-body attraction, interesting for some applications (see, for example, Ref.~\cite{Nishida}). Curiously, $g_3^{(2)}$ remains finite also for the ``magic'' angle conditioned by $\cos\theta = 1/\sqrt{3}$, a characteristic point where $V(x,{\bm y})$ looses its long-range dipolar tail in the $x$ direction.

As we have mentioned, for $\theta=0$ and $a=a_*$ the second-order term $g_3^{(2)}$ vanishes. The three-body interaction in this case emerges in the third order and can be repulsive. Evaluating Eq.~(\ref{g33}) and using Eq.~(\ref{Vnu0mu01D}) with $\theta=0$ and $a=a_*$, we obtain~\cite{RemExtrapolation}
\begin{equation}\label{g3mres}
g_3^{(3)}=4.65(r_*/l)^3.
\end{equation}
We see that the attractive long-range tail (corresponding to $r_*>0$) is correlated with a three-body repulsion. As in the quasi-two-dimensional case we can compare the full quasi-one-dimensional model with the purely one-dimensional one obtained by projecting the interaction potential $V({\bf r})$ to the radial ground state. The calculation of $g_3^{(3)}$ then proceeds by restricting the sum in Eq.~(\ref{g33}) to ${\bm \nu}={\bm \mu}={\bm \eta}=0$ and gives $g_3^{(3)}=3.57(r_*/l)^3$.

Finally, let us come back to the case of finite $\theta$ and mention the confinement-induced correction $g_2^{(2)}$. For $a=a_*$ Eq.~(\ref{g22}) contains no infrared divergences and the ultraviolet one has the same origin and is treated in the same manner as in the quasi-two-dimensional case. Performing a very similar analysis of the large-$s$ asymptote of $V_{{\bm 0} \{\mu,m'\} }^{{\bm 0} \{\nu,m\}}(k)$ we arrive at the diverging integral of the type (\ref{2ndOrderBorn}) for the function $V^2(k,{\bf p})$ with the prefactor $1/(2\pi l^2)$ instead of $1/(\sqrt{2\pi}l)$~\cite{remLocal}. When calculating Eq.~(\ref{g22}) we subtract this diverging contribution and arrive at
\begin{equation}\label{g2TiltedDipoles1D}
g_2^{(2)}=\frac{{r_*}^2}{l^3}(D_0+D_1\sin^2\theta+D_2\sin^4\theta)
\end{equation}
with the numerical coefficients $D_0=0.081$, $D_1=0.35$ and $D_2=-0.2$. The correction (\ref{g2TiltedDipoles1D}) is always positive. 

We can try to calculate $g_2^{(2)}$ by projecting to the transversal ground state, i.e., integrating over $k$ in Eq.~(\ref{g22}) with ${\bm \nu}={\bm \mu}=0$. In this manner we obtain $g_2^{(2)}=-0.94(r_*^2/l^3)[1-(3/2)\sin^2\theta]^2$, which is actually quite different from Eq.~(\ref{g2TiltedDipoles1D}), indicating that the correct renormalization procedure is important. In fact, Edler and co-workers~\cite{Edler2017} have calculated the BMF energy density for quasi-one-dimensional dipoles with $\theta=0$ and $a=a_*$ by using the projected value of $g_2^{(2)}$ as the low-density reference point for their Hugenholtz-Pines approach~\cite{LuisPrivate}. We agree with them on the effective three-body repulsion in this case, but disagree on $g_2$. This, however, does not qualitatively change the conclusion of Ref.~\cite{Edler2017} on the existence of self-bound states in this system since one can always tune $g_2$ by modifying $a$. Nevertheless, it would be interesting to perform the BMF crossover analysis using Eq.~(\ref{g2TiltedDipoles1D}) as the low-density reference point.

\section{Bogoliubov theory\label{Sec:Bogoliubov}}

The standard perturbation theory of Sec.~\ref{Sec:PT} requires that the interaction shifts be smaller than the level spacing in the non-interacting system of $N$ particles. In principle, one can always reach this regime by decreasing the amplitude of $V$ while keeping the volume fixed. The fixed volume maintains a low-momentum cutoff, avoiding possible infrared divergences in the integrals, and leads to the regular expansion of the energy in integer powers of $V$. 

The problem of infrared divergences can also be solved in the thermodynamic limit by turning to the Bogoliubov theory which accounts for a nonperturbative change in the system behavior at length scales comparable to the healing length $\xi \propto 1/\sqrt{V_{{\bm 0} {\bm 0}}^{{\bm 0} {\bm 0}}({\bf 0})n}$. The Bogoliubov theory thus effectively introduces an infrared density-dependent cutoff at $k\sim 1/\xi$, which, in particular, leads to the nonanalyticity of the energy as a function of $n$ (and $V$). 

In all examples of Sec.~\ref{app} the infrared divergences are eliminated by the condition (\ref{TheAssumption})~\cite{RemExotic}. Our theory thus predicts the regular expansion of the energy in integer powers of $n$, characterized by the effective few-body coupling constants $g_2$ and $g_3$, which are ``local'' BMF contributions involving virtual excitations with wave lengths comparable to the interaction range.

It has been noticed that the Bogoliubov theory also gives the regular density expansion under the assumption (\ref{TheAssumption})~\cite{Edler2017,Zin}. The reason is that the Bogoliubov vacuum is not perturbed as strongly as in the ``ordinary'' case of $V_{{\bm 0} {\bm 0}}^{{\bm 0} {\bm 0}}({\bf 0})\neq 0$. Note that the excitation spectrum at small momenta scales as $k\sqrt{nV_{{\bm 0} {\bm 0}}^{{\bm 0} {\bm 0}}(k)}$ and is softer than linear. Virtual processes at higher momenta and shorter length scales thus become relatively more important for the BMF term compared to what happens in the ``ordinary'' case. Low-momentum effects, like the roton or phonon instabilities, are expected to influence the system, but at a higher order. To illustrate this, the lowest Bogoliubov branch for quasi-two-dimensional dipoles with $\theta=0$ and $a=a_*$ equals $k^{3/2}\sqrt{k/4-r_*n}$ at $k\ll 1/l$ and features unstable modes for $k<4r_*n$ (we assume $r_*n\ll 1/l$). The energy density of the Bogoliubov vacuum is then characterized by an imaginary part $\propto (r_*n)^4$, which is much smaller that the two-body and three-body corrections calculated in Sec.~\ref{Sec:2Ddipoles}. Note also the separation of the characteristic momentum scales; putting the system in a box of linear size $L\sim 1/(r_*n)\gg l$ will hardly change the local few-body corrections but the unstable modes may disappear together with the corresponding imaginary part of the vacuum energy.  

The perturbation theory of Sec.~\ref{Sec:PT} is basically the Born series expansion for which the amplitude of the interaction potential (in real space) multiplied by the square of its range should be small. This small parameter is density independent. By contrast, the Bogoliubov theory relies on the relative smallness of the non-condensed fraction, which depends on the density. It is then interesting to figure out how the hierarchy of the Bogoliubov theory (the mean-field term, the leading-order BMF contribution, the beyond-Bogoliubov terms) is related to our expansion in powers of $V$. In the remaining part of this section we show how the perturbative results obtained in Sec.~\ref{Sec:PT} can be deduced from the Bogoliubov theory. 

The Hamiltonian (\ref{Ham}) in the second quantization reads
\begin{equation}\label{HamManyBody}
\begin{aligned}
&\hat{H}=\sum_{{\bm q},{\bm \nu}}(q^2/2+\epsilon_{\bm \nu})\hat{a}_{{\bm q},{\bm \nu}}^\dagger\hat{a}_{{\bm q},{\bm \nu}}\\
&+\frac{1}{2}\sum_{{\bm q}_1,{\bm q}_2,{\bm k},{\bm \nu},{\bm \mu},{\bm \eta},{\bm \zeta}}V_{{\bm \mu}{\bm \nu}}^{{\bm \zeta}{\bm \eta}}({\bf k})\hat{a}_{{\bm q}_2+{\bm k},{\bm \mu}}^\dagger\hat{a}_{{\bm q}_1-{\bm k},{\bm \zeta}}^\dagger\hat{a}_{{\bm q}_2,{\bm \nu}}\hat{a}_{{\bm q}_1,{\bm \eta}}.
\end{aligned}
\end{equation} 
Following the standard Bogoliubov procedure we assume the macroscopic occupation of the ground state replacing $\hat{a}_{{\bm 0},{\bm 0}}$ and $\hat{a}_{{\bm 0},{\bm 0}}^\dagger$ by $\sqrt{n_0}$ and then expanding $\hat{H}=H_0+\hat{H}_{sp}+\sum_{i=1}^4\hat{H}_i$, where
\begin{equation}\label{H0}
H_0=V_{{\bm 0}{\bm 0}}^{{\bm 0}{\bm 0}}({\bf 0})n_0^2/2,
\end{equation}
\begin{equation}\label{Hsp}
\hat{H}_{sp}=\sum_{{\bm q},{\bm \nu}}(q^2/2+\epsilon_{\bm \nu})\hat{a}_{{\bm q},{\bm \nu}}^\dagger\hat{a}_{{\bm q},{\bm \nu}},
\end{equation}
\begin{equation}\label{H1}
\hat{H}_1=n_0^{3/2}\sideset{}{'}\sum_{\bm \nu}V_{{\bm \nu}{\bm 0}}^{{\bm 0}{\bm 0}}({\bf 0})\hat{a}_{{\bm 0},{\bm \nu}}^\dagger+V_{{\bm 0}{\bm \nu}}^{{\bm 0}{\bm 0}}({\bf 0})\hat{a}_{{\bm 0},{\bm \nu}},
\end{equation}
\begin{equation}\label{H2}
\begin{aligned}
&\hat{H}_2=\frac{n_0}{2}\sideset{}{'}\sum_{{\bm \nu},{\bm \mu},{\bm k}}V_{{\bm \mu}{\bm 0}}^{{\bm \nu}{\bm 0}}({\bf k})\hat{a}_{{\bm k},{\bm \mu}}^\dagger\hat{a}_{-{\bm k},{\bm \nu}}^\dagger+V_{{\bm 0}{\bm \mu}}^{{\bm 0}{\bm \nu}}({\bf k})\hat{a}_{-{\bm k},{\bm \mu}}\hat{a}_{{\bm k},{\bm \nu}}\\
&\hspace{3cm}+2[V_{{\bm \mu}{\bm \nu}}^{{\bm 0}{\bm 0}}({\bf 0})+V_{{\bm \mu}{\bm 0}}^{{\bm 0}{\bm \nu}}({\bf k})]\hat{a}_{{\bm k},{\bm \mu}}^\dagger\hat{a}_{{\bm k},{\bm \nu}},
\end{aligned}
\end{equation}
\begin{equation}\label{H3}
\begin{aligned}
&\hat{H}_3=\sqrt{n_0}\sideset{}{'}\sum_{{\bm q},{\bm k},{\bm \nu},{\bm \mu},{\bm \eta}}V_{{\bm \nu}{\bm 0}}^{{\bm \mu}{\bm \eta}}({\bf k})\hat{a}_{{\bm k},{\bm \nu}}^\dagger\hat{a}_{{\bm q},{\bm \mu}}^\dagger\hat{a}_{{\bm q}+{\bf k},{\bm \eta}}\\
&\hspace{3cm}+V_{{\bm \eta}{\bm \mu}}^{{\bm 0}{\bm \nu}}({\bf k})\hat{a}_{{\bm q}+{\bf k},{\bm \eta}}^\dagger
\hat{a}_{{\bm k},{\bm \nu}}\hat{a}_{{\bm q},{\bm \mu}},
\end{aligned}
\end{equation}
and
\begin{equation}\label{H4}
\hat{H}_4=\frac{1}{2}\sideset{}{'}\sum_{{\bm q}_1,{\bm q}_2,{\bm k},{\bm \nu},{\bm \mu},{\bm \eta},{\bm \zeta}}V_{{\bm \mu}{\bm \nu}}^{{\bm \zeta}{\bm \eta}}({\bf k})\hat{a}_{{\bm q}_2+{\bm k},{\bm \mu}}^\dagger\hat{a}_{{\bm q}_1-{\bm k},{\bm \zeta}}^\dagger\hat{a}_{{\bm q}_2,{\bm \nu}}\hat{a}_{{\bm q}_1,{\bm \eta}}.
\end{equation} 
In Eqs.~(\ref{H1}-\ref{H4}) the primes indicate that the corresponding sum excludes terms involving creation or annihilation operators of condensate particles. 

Equation~(\ref{H0}) is the usual mean-field term. As far as the linear part Eq.~(\ref{H1}) is concerned, it appeared because we skipped one step of the standard Bogoliubov method. Namely, we just took the single-particle ground state $\phi_{{\bm 0},{\bm 0}}({\bf x},{\bf y})$ [see Eq.~(\ref{OneBodyEigenstates})] for the condensate mode instead of solving the mean-field Gross-Pitaevskii equation, which, in general, leads to a different profile in the confined direction (see, for example, \cite{Cai2010}). The inconvenience of having this linear term is compensated by the fact that the matrix elements $V_{{\bm \mu}{\bm \nu}}^{{\bm \zeta}{\bm \eta}}({\bf k})$ do not depend on the density and other parameters through the outcome of the Gross-Pitaevskii equation. In fact, this equation may not even always have a solution if the system is unstable from the mean-field viewpoint. It is also interesting to observe that tuning $V_{{\bm 0}{\bm 0}}^{{\bm 0}{\bm 0}}({\bf 0})$ to zero does not necessarily mean $\hat{H}_1=0$ since matrix elements of the type $V_{{\bm \nu}{\bm 0}}^{{\bm 0}{\bm 0}}({\bf 0})$ can still remain finite. A particular example of this phenomenon is quasi-one-dimensional dipoles with finite tilt discussed in Sec.~\ref{Sec:1Ddipoles}. In such cases, one can treat $\hat{H}_1$ as a perturbation on top of the single-particle Hamiltonian $\hat{H}_{sp}$ given by Eq.~(\ref{Hsp}). The second-order correction to the energy calculated in this manner equals $g_3^{(2)}n_0^3/6$, with $g_3^{(2)}$ given by Eq.~(\ref{g32}). 

In what follows we proceed under the assumption (\ref{TheAssumption}), which means $H_0=\hat{H}_1=0$. Among remaining terms the sum $\hat{H}_{sp}+\hat{H}_2$ is the quadratic Bogoliubov Hamiltonian, the zero-point energy of which gives the leading-order BMF contribution to the energy density. The diagonalization of this Hamiltonian consist of solving the linear Bogoliubov-de Gennes equations and can be done analytically in some cases (for instance, for flat condensates with periodic boundary conditions~\cite{Zin}). Otherwise, and this is the case of harmonic confinement, this procedure requires a numerical diagonalization of a $2M\times 2M$ matrix, where $M$ is the size of the discretized ${\bm \nu}$-space~\cite{Edler2017}. This is however, not necessary for our purposes. We just note that for small $V$ or $n_0$ we can treat $\hat{H}_2$ as a perturbation to $\hat{H}_{sp}$ and proceed with the standard perturbation theory. It is then easy to see that the first-order energy shift vanishes, and the second and third-order corrections are given by $g_2^{(2)}n_0^2/2$ and $g_3^{(3)}n_0^3/6$, respectively, where the coupling constants are given by Eqs.~(\ref{g22}) and (\ref{g33}). On the other hand, the third-order correction to the two-body constant Eq.~(\ref{g23}) is not recovered, which is understandable since $\hat{H}_2$ does not include interactions between excited atoms. 

In our search for all third-order corrections we thus have to formally go beyond the Bogoliubov approximation and consider $\hat{H}_3$ and $\hat{H}_4$. Note that these operators do not perturb the ground state of $\hat{H}_{sp}$ in any order. They can thus only react on the ground state already perturbed by $\hat{H}_2$ leading to corrections of order $V^3$ or higher. Indeed, the ground state of $\hat{H}_{sp}+\hat{H}_2$, calculated to the first order in $\hat{H}_2$, reads
\begin{equation}\label{Vaccum1}
\ket{1}=-\frac{n_0}{2}\sideset{}{'}\sum_{{\bm \nu},{\bm \mu},{\bm k}}\frac{V_{{\bm \mu}{\bm 0}}^{{\bm \nu}{\bm 0}}({\bf k})}{k^2+\epsilon_{\bm \nu}+\epsilon_{\bm \mu}}\hat{a}_{{\bm k},{\bm \mu}}^\dagger\hat{a}_{-{\bm k},{\bm \nu}}^\dagger\ket{0},
\end{equation}  
where $\ket{0}$ is the vacuum of excitations of $\hat{H}_{sp}$, i.e., pure condensate with density $n_0$. We then observe that $\bra{1}\hat{H}_3\ket{1}=0$ and the leading-order beyond-Bogoliubov contribution equals
\begin{equation}\label{H4aver}
\bra{1}\hat{H}_4\ket{1}=g_2^{(3)}n_0^2/2,
\end{equation} 
where $g_2^{(3)}$ is given by Eq.~(\ref{g23}). Finally, we note that the leading-order noncondensed density equals
\begin{equation}\label{Noncondensed}
\delta n = \bra{1}\sum_{{\bm q},{\bm \nu}}\hat{a}_{{\bm q},{\bm \nu}}^\dagger\hat{a}_{{\bm q},{\bm \nu}}\ket{1}=n_0^2\sum_{{\bf k},{\bm \nu},{\bm \mu}} \frac{|V^{{\bf 0 \bm \nu}}_{{\bf 0} {\bm \mu}}({\bf k})|^2}{(k^2 + \epsilon_{\bm \nu} + \epsilon_{\bm \mu})^2},
\end{equation} 
and, to the order $V^3$, for all the above mentioned energy corrections we can take $n_0$ to be equal to the total density. We have thus established the consistency of the second-quantized Bogoliubov approach with the standard first-quantized approach of Sec.~\ref{Sec:PT} up to the order $V^3$. Note, however, that Eq.~(\ref{Noncondensed}) features an infrared logarithmic divergence for quasi-two-dimensional dipoles since $V^{{\bf 0 \bm \nu}}_{{\bf 0} {\bm \mu}}({\bf k})\propto k$. This divergence leads to the scaling $\delta n\propto V^2\ln V$, which does not change our conclusion, but signals that at higher orders a nonperturbative treatment of the quadratic Bogoliubov Hamiltonian is necessary in this case.

\section{Conclusions}\label{Sec:Conclusions}

In conclusion, we have developed a perturbation approach for calculating interaction energy shifts for bosons with the interaction potential $V$ tuned close to the condition (\ref{TheAssumption}). Under this assumption the leading nonpairwise energy correction is a third-order effect manifesting itself in the form of an effective three-body interaction. Whether this interaction is attractive or repulsive is determined by the shape of $V({\bf k})$ through Eq.~(\ref{g33}). However, for minimally exotic potentials [for instance, for those featuring no nodes in $V({\bf k})$], the sign of $g_3$ is inversely correlated with the sign of the long-range tail of $V({\bf r})$. 

We have applied our theory to a few particular shapes of $V$ in pure dimensions (double-Gaussian and Yukawa-plus-delta potentials) and in quasi-low-dimensional geometries where we have considered tilted dipoles. For the latter systems we have fully characterized the leading two-body and three-body energy corrections as a function of the tilt angle. In particular, we have found that dipoles under harmonic quasi-two-dimensional confinement are characterized by an effective three-body attraction when aligned perpendicularly to the plane and by a three-body repulsion if aligned in the plane (see Fig.~\ref{Fig:g3}). It remains to see if this repulsion can stabilize dilute supersolid stripe phases of tilted dipoles, so far predicted to be stable in the dense regime~\cite{Macia2012,Macia2014}). The three-body repulsion for dipoles aligned in the plane has also been found by Zin {\it et al.}~\cite{Zin}, although in their case the quasi-two-dimensional confinement is achieved by imposing the periodic boundary condition. 

Our analysis of quasi-one-dimensional dipoles has revealed a strong (second-order) three-body attraction for any finite tilt angle and a weaker (third-order) three-body repulsion when the dipoles are aligned along the unconfined axis. This latter observation is in agreement with the BMF calculations of Ref.~\cite{Edler2017}. We however disagree on the leading-order two-body correction $g_2^{(2)}$, positive in our case and negative in Ref.~\cite{Edler2017} (see our comment in the end of Sec.~\ref{Sec:1Ddipoles}). We argue that our results can be used to improve the Hugenholtz-Pines analysis by providing the low-density reference point for quasi-low-dimensional tilted dipoles.   

\section*{Acknowledgements} We acknowledge support from ANR Grant Droplets No. ANR-19-CE30-0003-02.


\begin{thebibliography}{99}
\bibitem{RemStat}
Distinguishable atoms have the same ground state as identical bosons.
\bibitem{RemExtrapolation}
Introducing a cutoff $\epsilon_c \propto 1/r_0^2$ for excitation energies in Eq.~(\ref{g33}) we clearly observe the scaling $g_3(\epsilon_c)-g_3(\infty)\propto 1/\sqrt{\epsilon_c}$. We use this scaling for calculating $C_0$, $C_1$ in Eq.~(\ref{g3TiltedDipoles}) and the numerical coefficient in Eq.~(\ref{g3mres}).
\bibitem{RemExotic} There are, of course, examples featuring infrared divergences on the one hand, and satisfying Eq.~(\ref{TheAssumption}) on the other. Consider, for instance, the low-momentum scaling $V(k)\propto |k|^\gamma$ with small positive $\gamma$.
\end{thebibliography}
\end{document}